\documentclass{article}

\usepackage{arxiv}

\usepackage[utf8]{inputenc} 
\usepackage[T1]{fontenc}    
\usepackage{hyperref}       
\usepackage{url}            
\usepackage{booktabs}       
\usepackage{amsfonts}       
\usepackage{nicefrac}       
\usepackage{microtype}      
\usepackage{lipsum}
\usepackage{graphicx}
\usepackage{verbatim}
\graphicspath{ {./figures/} }
\usepackage{subfig}
\usepackage{xcolor}
\usepackage{setspace}
\usepackage{url}

\title{Controversial information spreads faster and further in Reddit}

\author{
 Jasser Jasser \\
  Department of Computer Science\\
  University of Central Florida\\
  Orlando, FL, United States \\
  \texttt{Jasser.Jasser@ucf.edu} \\
   \And
 Ivan Garibay \\
  Department of of Industrial Engineering and Management Systems\\
  University of Central Florida\\
  Orlando, FL, United States \\
  \texttt{igaribay@ucf.edu} \\
  \And
 Steve Scheinert \\
  Deloitte Consulting, LLC\\
  Lake Mary, FL, United States\\
  \And
 Alexander V. Mantzaris \\
 Department of Statistics and Data Science \\
 University of Central Florida \\
 Orlando, FL, United States
}

\begin{document}
\maketitle
\begin{abstract}
Online users discuss and converse about all sorts of topics on social networks. Facebook, Twitter, Reddit are among many other networks where users can have this freedom of information sharing. The abundance of information shared over these networks makes them an attractive area for investigating all aspects of human behavior on information dissemination. Among the many interesting behaviors, controversiality within social cascades is of high interest to us. It is known that controversiality is bound to happen within online discussions. The online social network platform Reddit has the feature to tag comments as controversial if the users have mixed opinions about that comment. The difference between this study and previous attempts at understanding controversiality on social networks is that we do not investigate topics that are known to be controversial. On the contrary, we examine typical cascades with comments that the readers deemed to be controversial concerning the matter discussed. This work asks whether controversially initiated information cascades have distinctive characteristics than those not controversial in Reddit. We used data collected from Reddit consisting of around 17 million posts and their corresponding comments related to cybersecurity issues to answer these emerging questions. From the comparative analyses conducted, controversial content travels faster and further from its origin. Understanding this phenomenon would shed light on how users or organization might use it to their help in controlling and spreading a specific beneficiary message.
\end{abstract}

\keywords{Controversiality \and Information \and Diffusion \and Reddit \and Polarization}

\section{Introduction}
\label{intro}
The emergence of online social networks opened the door for many to amplify content freely. Much of this activity is benign or even positive. However, this openness formed a platform for the polarization of opinions and controversial discussions. There is active research into polarization on online social networks \cite{taylor2018exploring,akoglu2014quantifying,garimella2018quantifying}. Controversiality of content is one feature that could draw attention to that content \cite{kim1999news}. The spreading and promotion of controversial content support the polarization of social networks, especially in the political sphere \cite{conover2011political}. A question naturally arises of whether such behavior is beneficial to the content's promoter and that there is an incentive for the activity. This work investigates this question and whether controversy within a discussion would bring more user attention to authored content and promote the spread of the material further and faster.

Reddit is a popular online platform that allows users to spread knowledge and share opinions through posts consisting of both textual and visual elements. It allows users to respond to others' posts by either commenting or voting for or against other users' posts and comments, known as up- and down-votes  \cite{stoddard2015popularity}. Social media users display a tendency to join homogeneous communities that share the same beliefs and are dedicated to the same topic (echo chambers \cite{sunstein2001echo,garrett2009echo}). Reddit provides spaces, known as \textit{subreddits}, where these discussions occur and which facilitate more focused discussion, and the presence of controversiality is genuine and not artificial. A study by the Pew Research Center (2013) concluded that 6\% of online adults are Reddit users \cite{duggan20136}, and by the time of writing, Alexa\footnote{A web traffic analysis company owned by Amazon} ranks it as \#6 in the United States and \#15 globally\footnote{https://www.alexa.com/siteinfo/reddit.com}.Reddit uses up- and down-votes to identify controversial content;  with these votes representing agreement or disagreement, respectively, if the comment received a fair amount of polarized votes, then the comment is considered controversial. The identification of controversial content is determined by the Reddit's formula for controversial comments, the final labeling for which is provided through the Reddit API. According to Reddit's definition, a comment is controversial if 1) the sum of up- and down-votes is greater than or equal to 7, and 2) the ratio of up- and down-votes lies between selected upper ($ub$ = 0.6) and lower bounds ($lb$ = 0.4) as shown in equation (\ref{eq:reddit-controversial}). We found the algorithm responsible for identifying controversial
comments in the \textit{reddit-archive} repository in GitHub\footnote{ https://github.com/reddit-archive/reddit/blob/master/r2/r2/models/builder.py}.
\begin{equation}
    C_i = \left\{
                \begin{array}{ll}
                  True, \quad lb < c < ub\\
                  False, \quad else\\
                \end{array}
              \right.
\label{eq:reddit-controversial}
\end{equation}
Here $C_i$ is the comment index and whether it is classified as controversial or not. Additionally, \textit{subreddit} moderators may tag a post or comment as controversial, regardless of the number of votes. We assume that the tagging of a comment to be controversial is the result of users being invested in the discussion and that the polarization in the voting comes as users are swayed by some aspects of the discussion.

We rely on the definition and labeling of controversiality that Reddit applies to comments made on the platform. We understand the label to mean that there was attention to what is being discussed and there was a polarized reaction to the labeled comment. The measure of the effect of the resurgence of such labeled comments is studied by analyzing the characteristics of the posts' cascades that contain these comments and those that do not, and by investigating the independent authors' networks that forms these posts' cascades. The results show that conversations that contain controversial content during the early stages of their lifetime possess more activity, and users involved in that conversation have a wider influence than conversations that are not labeled as controversial.

\section{Related Work}
The concept of polarization and controversiality has been explored in multiple fields within the literature, from its identification in social networks and blogs \cite{Adamic,conover2011political,An} to its detection in online news and web pages \cite{Kaplun,choi2010identifying,Dori-Hacohen}. There are two main themes of the research, the \textit{identification} and \textit{quantification} of controversy within the discussion. Other literature explores the effect of controversial information such as the work of  \cite{garimella2018quantifying}, which studies the effect of collective attention on the evolution of controversial debates. The work of \cite{mejova2014controversy} explores the effect of the controversy on the emotions and language within online news. 

The identification of controversy in social discussions is an area of on-going research with many different approaches to the identification of controversiality. The work of \cite{choi2010identifying} identifies controversy within a topic by looking at the sentiment of the discussion and the range of sentiment polarity across the users involved in the controversy, as measured through natural language processing and sentiment analysis. Similarly in the work of \cite{Popescu}, the authors proposed identifying controversial events by also considering the measurement of polarized opinions in the wake of a widely-viewed event. The work of \cite{qiu2019investigating} investigated opinion distribution in social media to identify controversial discussions and the size of their opinion groups.

Case studies that have attempted to quantify controversy and controversiality have focused on techniques and methods to measure the amount of controversy within a network or discussion. The work of \cite{garimella2018quantifying} quantifies controversy in online echo chambers by considering a topic that is controversial and building the conversation graph of that topic which depicts the opinion alignment between the users. The authors then partition the graph to identify the sides of the controversy where the amount of controversy is measured from the characteristics of that graph. The work of \cite{guerra2013measure}, compares two types of social networks, those who are formed from polarized context and those who are not by analyzing the boundary between a pair of communities. The authors then distinguish between the polarized from the non-polarized communities by the concentration of high-degree nodes found in the boundary. The authors found that polarized networks tend to have low concentration "popular (high-degree)" nodes along the boundary between communities. 

\section{Data and Investigation  }

\begin{figure*}[ht]
\centering
\subfloat[]{\includegraphics[width = 0.3\linewidth]{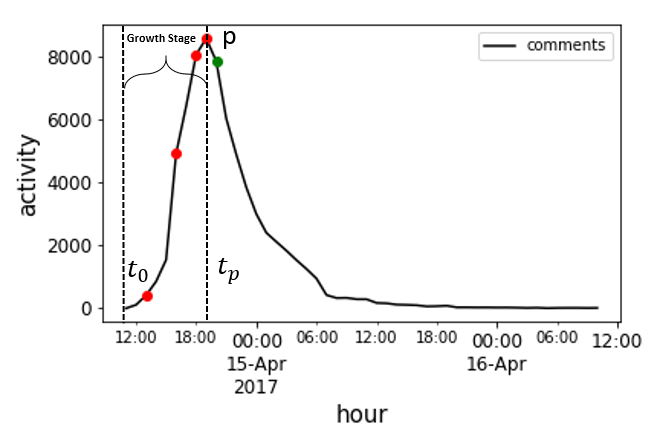}} 
\subfloat[]{\includegraphics[width = 0.3\linewidth]{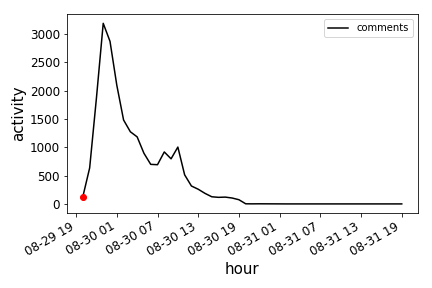}}
\subfloat[]{\includegraphics[width = 0.3\linewidth]{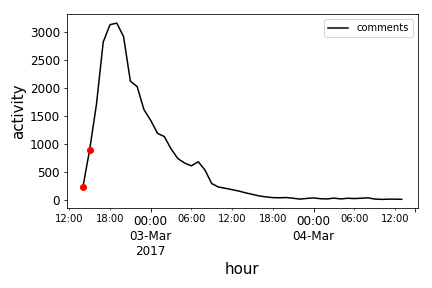}} \\
\subfloat[]{\includegraphics[width = 0.3\linewidth]{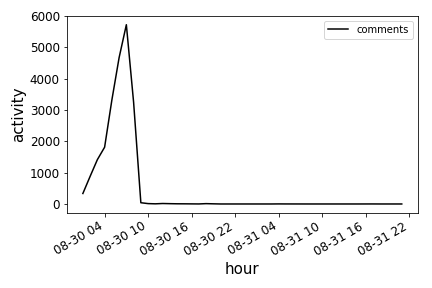}} 
\subfloat[]{\includegraphics[width = 0.3\linewidth]{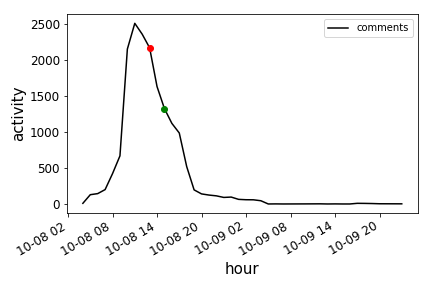}}
\subfloat[]{\includegraphics[width = 0.3\linewidth]{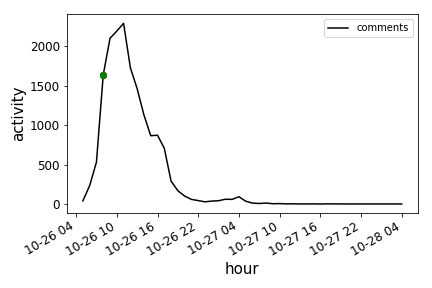}} 
\caption{Controversially initiated and non-controversially initiated cascades, (a,b,c) are controversially initiated posts' cascades while (d,e,f) are non-controversial posts' cascades where the red dots represent a comment labeled as controversial by Reddit that is directed to the post's author while a green dot is a comment labeled controversial by Reddit that is directed to another comment.}
\label{fig:controversial-non-controversial}
\end{figure*}

The data used for this investigation is collected from Reddit and provided as part of the "Computational Simulation of Online Social Behavior (SocialSim)" DARPA program\footnote{https://www.darpa.mil/program/computational-simulation-of-online-social-behavior}. This vast dataset consists of more than 36 million comments on posts related to cybersecurity issues. For this study, we considered the posts in our dataset that have at least 100 comment. The number of posts collected is 47,940 where the total comments in those posts exceeded 17 million comments. the posts are then analyzed to be classified either as controversially initiated or non-controversial based on whether there exist any comments that are labeled controversial according to Reddit (as explained above) within the early stages of the posts. The classification yielded 23,101 posts that are labeled as controversially initiated, and 24,839 are labeled non-controversial. The distribution of controversial and non-controversial cascades from the same \textit{subreddits} is almost equal in quantity. As an example, there are 8,528 controversial and 12,349 non-controversial cascades coming from the ‘\textit{pcmasterrace}’ \textit{subreddit}, and 5,450 controversial and 4,149 non-controversial cascades coming from the ‘\textit{android}’ \textit{subreddit}.


In social networks such as Reddit, users publish posts in regard of a specific topic where they share information and opinions. The posts are time stamped and so are the comments from the users that replied to that post. This forms a temporal propagation graph, also known as information cascade \cite{bikhchandani1992theory,leskovec2007cost}. The classification of cascades is based on the enclosure of controversially labeled comments within the posts that form the cascade. Cascades have a growth stage which is the period between the time of the first comment ($t_0$) and the peak time ($t_p$) of the post's cascade as shown in Fig\ref{fig:controversial-non-controversial}(a). We define a cascade that follows a post as controversial if the cascade contains a comment, directed to the post's author, that is labeled controversial and occurred within the growth stage. Assuming that the discussion reached a level of polarity in the opinions within the growth stage of the cascade that drove the discussion. We do not consider controversial comments that happened after the growth stage since at that stage users have lost interest in the topic and the polarization did not reinvigorate the discussion. Fig\ref{fig:controversial-non-controversial} shows posts' cascades where Subfigures \ref{fig:controversial-non-controversial}(a,b,c) are classified as controversial while Subfigures \ref{fig:controversial-non-controversial}(d,e,f) are classified as non-controversial. Notice Subfigures \ref{fig:controversial-non-controversial}(e,f) where in Subfigure \ref{fig:controversial-non-controversial}(e) there are two controversial comments after its growth stage and for that it was classified as non-controversial, and in Subfigure \ref{fig:controversial-non-controversial}(f) there is one controversial comment that is within the growth stage, however, it was not directed to the post's author and so it was not classified as controversial.

In our assumption, when the Reddit algorithm labels a comment as controversial, it is a sign of collective attention \cite{Wu17599} where users taking part in that discussion are still active and paying attention. The attention in large groups to the topics indicates popularity, where the popular the topic, the faster and further the information sharing and dissemination \cite{Wu17599} The comments directed to the post author are more visible to the reader, thus poised to grab more attention, and for that, we prioritized them in our analyses. As an example, the largest cascade in our dataset (controversially initialized) is about a giveaway for a personal computer. Comments directed to the post’s author such as “Give it to meeeh Im poor,” and “Please for the love of God please pick me I never win anything” got attention with polarized responses and were labeled controversial by Reddit’s algorithm. The attention within this cascade drove it to end up having around 60,000 comments. While the largest non-controversial cascade was also concerning a personal computer giveaway, but it did not hold any controversially labeled comments directed to the post’s author and did not achieve the high amount of comments (around 23,000 comments) mentioned in the previous cascade. However, there are controversially labeled comments that are not directed to the post’s author and occurred after the growth stage of the cascade where the attention and discussion have decayed.  

\begin{figure}[ht]
\centering
\subfloat[]{\includegraphics[width = 0.4\linewidth]{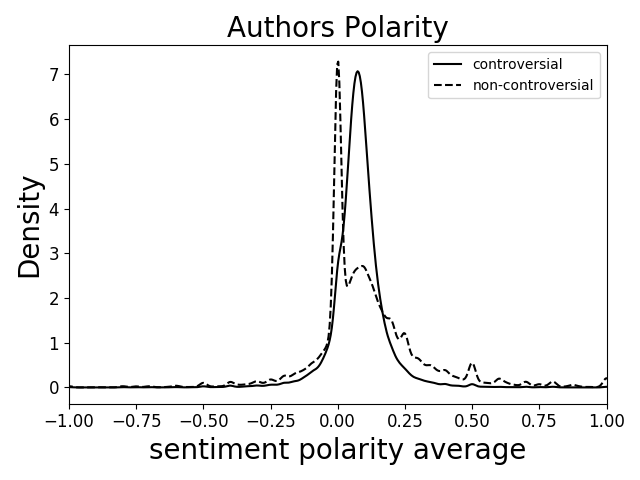}}
\subfloat[]{\includegraphics[width = 0.4\linewidth]{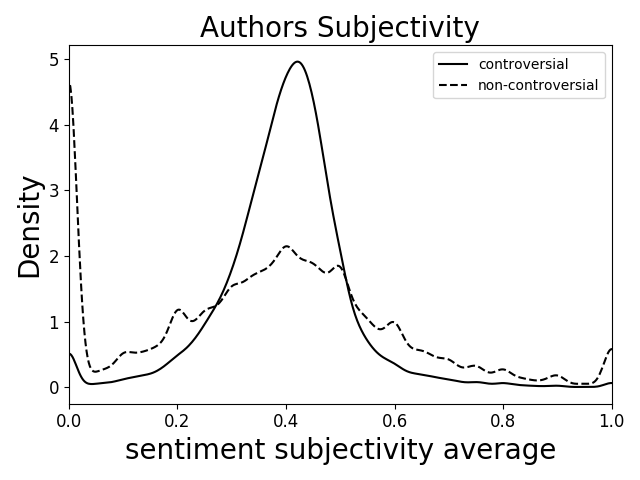}}
\caption{Authors' averages of sentiment polarity and subjectivity, (a) shows the authors' average for sentiment polarity while (b) shows the authors' average for sentiment subjectivity }
\label{fig:controversial_authors}
\end{figure}

The data come with the sentiment polarity and subjectivity of the comments calculated using the "pattern.en" model\footnote{https://www.clips.uantwerpen.be/pages/pattern-en}. The model ranks the polarity of the comment sentiment based on the contained adjectives on a scale from -1 to 1 where -1 means highly negative sentiment while 1 is a highly positive sentiment. The subjectivity is ranked by the model on a scale from 0 to 1 where 0 is very objective and 1 is very subjective. The sentiment polarity and subjectivity of the authors were calculated by taking the average sentiment of all comments written by that author. If an author wrote a comment that got labeled as controversial by Reddit, that author is labeled as controversial. The dataset contains  1,172,886 unique authors. Controversial authors consist of around 0.06\% of that total or 69,607 authors to be precise.
Figure \ref{fig:controversial_authors} shows that authors that are labeled controversial tend to have balanced polarity that is normally distributed with a mean slightly above 0, meaning they are more likely to be positive in their comments. 
Unlike the controversial authors, non-controversial authors mostly have no sort of polarity in their text. putting a spike in their distribution at 0. However, like controversial authors, this is a smaller, second peak at slightly positive, suggesting a bi-modal distribution. 
In terms of subjectivity, controversial authors tend to have a balance between being objective or subjective about the topic, hence their sentiment subjectivity average are closer to 0.5, while non-controversial authors are mostly objective in their comments. 

A social network is a form of complex network \cite{easley2010networks}, a social structure that consists of social actors and the relationships between them. The relationship and connectedness between the actors define and distinguish those that are influencing the network from those who do not. The work of \cite{kitsak2010identification}, argues that most influencers are found within the $k$-core of the network. A $k$-core is the "largest subgraph where vertices have at least $k$ interconnections" \cite{dorogovtsev2006k}. We considered the independent authors that commented on each other comments for every post to measure the connectedness among them and discovering the subset of nodes with the highest \textit{coreness}. This subset of authors defines the nodes responsible for influencing the dissemination of information. The $k$-core of that graph network is analyzed where $k$ is $max(coreness)$. We are interested in finding the number of nodes that are labeled controversial and their ratio within these $k$-cores. 


We investigated whether the earlier the resurgence of the controversial comments within the cascade would lead to drastic changes in the characteristics by defining epochs within the growth stage of the cascade. The period is then divided into three epochs, epoch 1 for the first quartile, epoch 2 for the second quartile, and epoch 3 for the second half of the growth stage. If one of these epochs contains the majority of the controversial comments, we consider that epoch as the period with the highest concentration of controversy, and we associate the post's cascade with that period. An epoch 1 cascade means that the cascade's highest concentration of controversiality happened during the first quartile of the growth stage.

\section{Results}

\begin{figure}[ht]
\centering
\subfloat[]{\includegraphics[width = 0.4\linewidth]{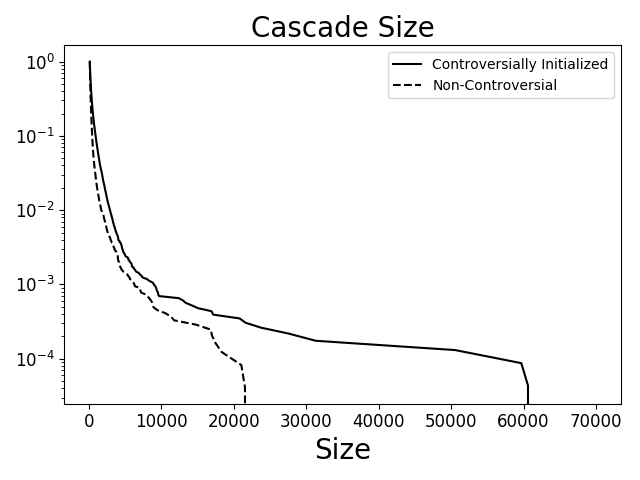}} 
\subfloat[]{\includegraphics[width = 0.4\linewidth]{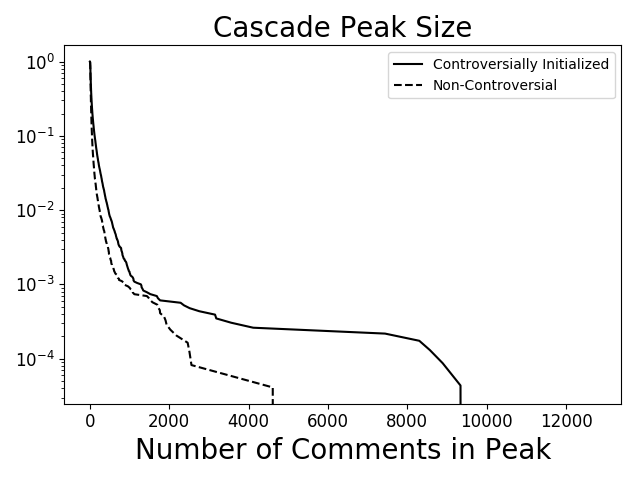}} \\
\subfloat[]{\includegraphics[width = 0.4\linewidth]{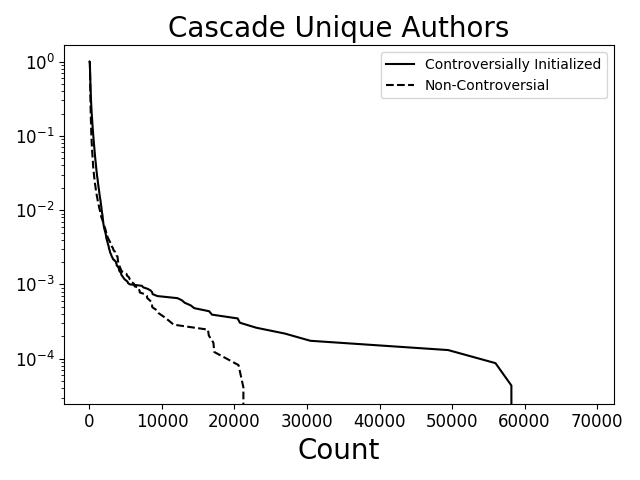}}
\caption{Descriptive analysis results of posts' cascade, (a) shows the difference in size between controversially initiated post's cascade and non-controversial one. (b) shows the difference in peak size, and (c) shows the difference in the number of unique authors}
\label{fig:ccdfs-cascades}
\end{figure}

The results of the analysis show that posts' cascades that contained controversially labeled messages during its growth stage produce a larger total number of comments and a larger burst, indicating more involvement from users. Another feature is that more unique users are involved in the cascades which is controversially labeled. This clearly shows that its beneficial for a user who seeks to spread a particular message faster and further, to infuse controversiality within the context of the message. The plots in Figure \ref{fig:ccdfs-cascades} show this effect presented on three different measurements. Each plot is a CCDF (complementary cumulative distribution function) where the x-axis is the quantity in question shown in the title and the y-axis is the probability for that measure to exceed that x-axis value. In subfigure \ref{fig:ccdfs-cascades}(a), shows that the posts that attract the largest contributions are dominated by initial signals of controversy. As the cascades grew, a difference emerged; non-controversial cascades did not exceed 25,000 comments whereas the controversial cascades did display posts having 60,000 or more comments. Subfigure \ref{fig:ccdfs-cascades}(b) shows the peak size probability  where controversial cascades tend to generally have larger peaks. Subfigure \ref{fig:ccdfs-cascades}(c) shows the probability for cascades to exceed a certain number of unique authors contributing comments to a post. Along with this measure it can be seen how the controversially initiated cascades produce much larger values for the upper tail of the distribution.

\begin{figure}[ht]
\centering
\subfloat[]{\includegraphics[width = 0.4\linewidth]{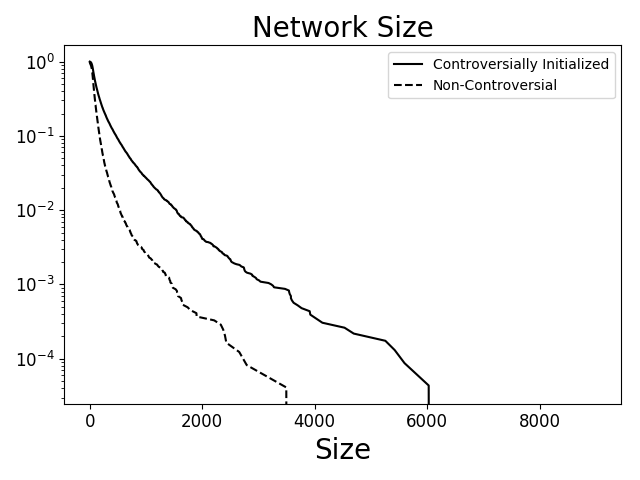}} 
\subfloat[]{\includegraphics[width = 0.4\linewidth]{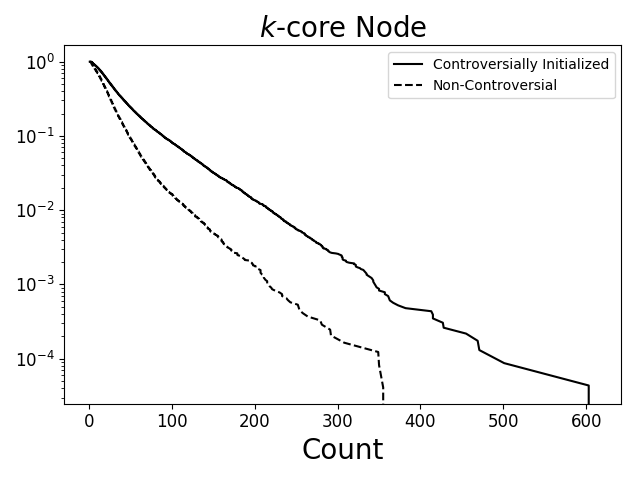}} \\
\subfloat[]{\includegraphics[width = 0.4\linewidth]{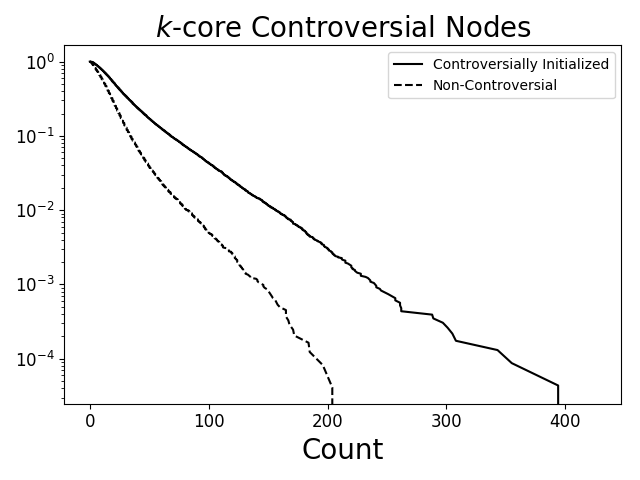}}
\caption{Network analysis results of posts' cascade, (a) shows the difference in network size (number of edges) between controversially initiated post's cascade and non-controversial one. (b) shows the difference in the number of nodes within the networks' $k$-core, and (c) shows the difference in the number of controversial nodes within the networks' $k$-core }
\label{fig:ccdfs-network}
\end{figure}

Figure \ref{fig:ccdfs-network} shows the results from conducting a network analysis on the post structure in the Reddit dataset. The results here show a more distinctive differentiation between the two types of cascades. Subfigure \ref{fig:ccdfs-network}(a) shows the CCDF for the network size of the controversially and non-controversially initiated cascades where the network size is computed by using the total number of edges produced within the entity of the post. Subfigure \ref{fig:ccdfs-network}(b) looks at the number of nodes within the defined $k$-core in the cascades' network. It is clearly visible that controversial cascades end up containing more nodes within their $k$-core and so are the controversial nodes within these $k$-cores as shown in Subfigure \ref{fig:ccdfs-network}(c). This shows that network properties of the different cascades are more affected by content which attracts controversial initiations or that they shape the subsequent discussions.

\begin{figure}[ht]
\centering
\subfloat[]{\includegraphics[width = 0.4\linewidth]{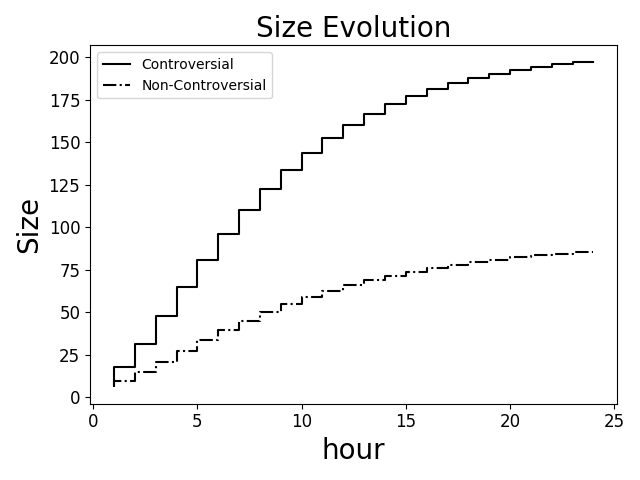}} 
\subfloat[]{\includegraphics[width = 0.4\linewidth]{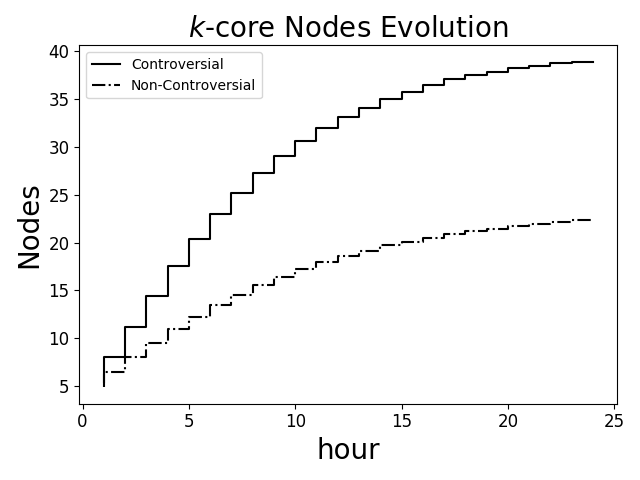}} \\
\subfloat[]{\includegraphics[width = 0.4\linewidth]{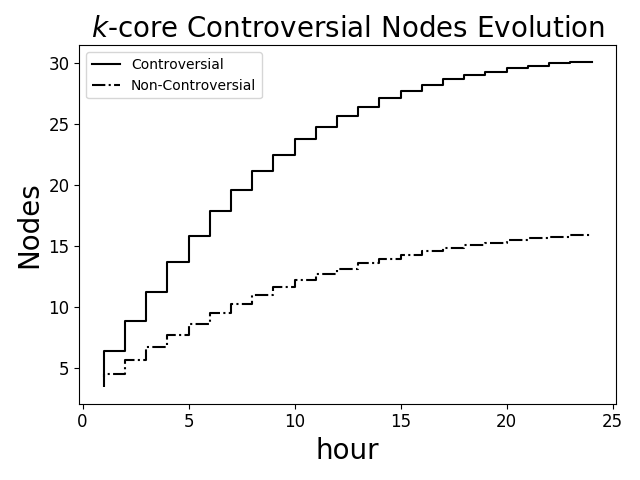}}
\caption{Network evolution analysis results of posts' cascade, (a) shows the difference in network size (number of edges) hourly growth between controversially initiated post's cascade and non-controversial one. (b) shows the difference in the hourly growth of the number of nodes within the networks' $k$-core, and (c) shows the difference in the hourly growth of the number of controversial nodes within the networks' $k$-core }
\label{fig:networkevo}
\end{figure}

Figure \ref{fig:networkevo} shows an evolutional temporal analysis of the network changes within the first 24 hours. Subfigure \ref{fig:networkevo}(a) shows the evolution the number of links in the network. The activity for the controversial cascades is greater and for the time duration observed has this increase from each in the post creation. Subfigure \ref{fig:networkevo}(b) shows the $k$-core evolution where the number of nodes within the controversial $k$-core increased drastically over the other ones. Subfigure \ref{fig:networkevo}(c) compares the evolutional number of controversial nodes found within the $k$-core. This reinforces that the cascades which contain controversial activity during their early stages continue to swiftly progress for longer periods of time than the non-controversial cascades.

\begin{figure}[ht]
\centering
\subfloat[]{\includegraphics[width = 0.4\linewidth]{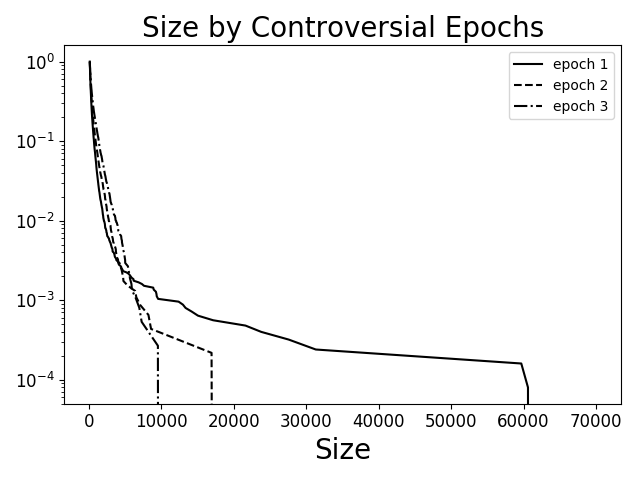}}
\subfloat[]{\includegraphics[width = 0.4\linewidth]{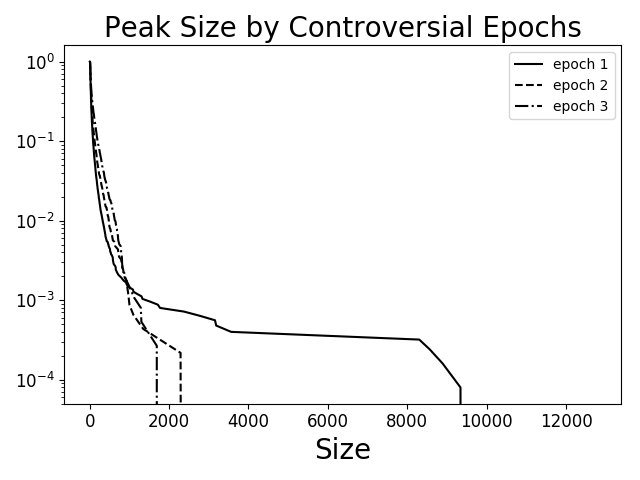}}
\caption{The posts' cascades size and peaks size based on the epoch that contains the highest concentration of the controversial comments. (a) shows the comparison of posts' cascades size, while (b) shows the comparison of the peak size}
\label{fig:periods}
\end{figure}

 Figure\ref{fig:periods} shows the results of comparing the first three epochs by the cascade size and the peak size of the cascade. Subfigure \ref{fig:periods}(a) shows that epoch 1 controversial cascades produced the largest number of comments and peak size, followed by epoch 2 then epoch 3 cascades. This indicates that the earlier the resurgence of the controversiality within the cascade the further the information will disseminate within, while later resurgence of controversial comments might not drastically affect the information flow.

\section{Conclusion}

The work sought to explore whether controversial content will have a greater chance of  increased activity amongst other users. The analysis uses descriptive statistics and network analysis for the amount of activity a cascade produced. The $k$-core analysis conducted to study the structure of the networks' core. The temporal dimension is also explored in an analysis of the evolution of contributors. The results show that content which was controversial is associated with higher degrees of activity. This can shed light on various marketing strategies for competing for attention online. Even if not completely understood as to why this occurs, it can be utilized by accounts on social media seeking to create more influence. It can also help explain the spread of disinformation. 

\section*{acknowledgements}
{Acknowledgment}
This work was partially supported by grant FA8650-18-C-7823 from the Defense Advanced Research Projects Agency (DARPA). The views and opinions contained in this article are the authors and should not be construed as official or as reflecting the views of the University of Central Florida, DARPA, or the U.S. Department of Defense.


\end{document}